\documentclass[a4paper]{ISOStyle}
\usepackage{epsfig}

\newcommand{\Lsun}{L$_{\odot}$}
\newcommand{\Msun}{M$_{\odot}$}
\newcommand{\Rsun}{R$_{\odot}$}
\newcommand{\cc}{$\mathrm{^{12}C/^{13}C}$}
\newcommand{\ad}{$\theta_d$}
\newcommand{\vt}{$\xi_t$}

\newcommand{\km}{$\rm{km\,s^{-1}}$}

\newcommand{\teff}{T$_{\mathrm{eff}}$}

\newcommand{\mic}{$\mu$m}
\newcommand{\be}{\begin{equation}}
\newcommand{\ee}{\end{equation}}
\newcommand{\bee}{\begin{eqnarray}}
\newcommand{\ene}{\end{eqnarray}}

\begin{document}

\title{ISO IMPACT ON STELLAR MODELS AND VICE VERSA}

\author{L. Decin\inst{1} \and C. Waelkens\inst{1} \and K. Eriksson\inst{2} \and
  B. Gustafsson\inst{2} \and B. Plez\inst{3} \and A.J. Sauval\inst{4} \and
  W. Van Assche\inst{5} \and B. Vandenbussche\inst{1}} \institute{Instituut voor
  Sterrenkunde, KULeuven, Celestijnenlaan 200B, B-3001 Leuven, Belgium \and
  Astronomiska Observatoriet, Box 515, S-75120 Uppsala, Sweden \and GRAAL -
  CC72, Universit\'{e} de Montpellier II, 34095 Montpellier Cedex 5, France \and
  Observatoire Royal de Belgique, Avenue Circulaire 3, B-1180 Bruxelles, Belgium
  \and Instituut voor Wiskunde, KULeuven, Celestijnenlaan 200B, B-3001 Leuven,
  Belgium}

\maketitle 

\begin{abstract}
We present a detailed spectroscopic study of a sample of bright, mostly cool,
stars observed with the Short-Wave\-length Spectrometer (SWS) on board  ISO,
which enables the accurate determination of the stellar parameters of the cool
giants, but also serves as a critical review of the ISO-SWS calibration.
\vspace*{-.2truecm}
\end{abstract}

\vspace*{-.7cm}
\section{INTRODUCTION}

ISO has opened the infrared window for detailed spectroscopic analysis. The
modeling and interpretation of the ISO-SWS data requires an accurate calibration
of the spectrometers (\cite{DWE_sch96}). In the SWS spectral region (2.38 - 45.2
\mic) the primary standard calibration sources are bright, mostly cool, stars.

It is obvious that the calibration of ISO cannot be more accurate
than our understanding of the observations of standard stars.  But
precisely because spectroscopic observations with this resolution
were not possible before this infrared window was first covered by
ISO, our understanding of stellar sources -and more precisely
stellar atmospheres- is not as refined as in the spectral range
that is accessible to ground-based instruments. A full
exploitation of the ISO data will therefore result from an
iterative process, in which both accurate observations and new
modeling are involved.  

To generate the stellar atmospheric models and the
synthetic spectra the MARCS-code (\cite{DWE_gust75}, \cite{DWE_plez92},
\cite{DWE_plez92}) was 
used. So far, the analysis of the discrepancies between the ISO-SWS data 
and the corresponding synthetic spectra has been restricted to the
wavelength region from 2.38 to 12 \mic, since the lack of
extensive molecular and atomic line lists hamper fast progress at
longer wavelengths (12 - 45 \mic). Furthermore the brightness of
the stars drops quickly in this wavelength region so that the same
signal to noise ratio will not be achieved. A third point is that
the spectral energy distributions (SEDs) may also be affected by
unknown circumstellar contributions.

Precisely because this research involves both theoretical developments on
the model spectra and calibration improvements on the spectral reduction, one
has to be extremely careful not to confuse technical detector problems with
astrophysical issues. Therefore several precautions are taken. They are
elaborated on in the next section.

\vspace*{-.3truecm}
\section{METHOD OF ANALYSIS}\label{DWE_sec:meth}

\subsection{Selection criterion}

Stellar standard candles spanning the spectral types A0-M8 were observed in the
framework of this research. It is important to cover a broad parameter space in
order to distinguish between calibration problems and problems related to the
model and/or the generation of the synthetic spectrum. Stars cooler than an M2
giant -i.e. cooler than 
$\sim 1500$~K- have not been scrutinized carefully. Calibration problems with
\object{$\gamma$ Cru}, variability (\cite{DWE_mon98}), the possible presence of
a circumstellar envelope, stellar winds or a warm molecular envelope above the
photosphere (\cite{DWE_tsuji97}) made the use of hydrostatic models for these
stars implausible and have led to the decision to postpone the modeling of the
coolest stars in the sample. 
From now on, we will distinguish between {\it{hot}} and
{\it{cool}} stars in the following way. {\it{Hot}} stars are stars hotter than
the Sun and their spectra are mainly
dominated by atomic lines, while the spectral signature of {\it{cool}} stars is
dominated by molecules.

\vspace*{-.3truecm}
\subsection{Data reduction}
In order to reveal calibration problems, the ISO-SWS data have to
be reduced in a homogeneous way. For all the stars in our
sample, at least one AOT01 observation is available, some stars
have also been observed using the AOT06 mode.  Since these AOT01
observations form a complete and consistent set, they were used as
basis for the research. In order to check potential calibration
problems, the AOT06 data are used. In this way both integrity and
security are implemented. The scanner speed of the AOT01
observations was 3 or 4, resulting in a resolving power $\simeq$
870 or $\simeq$ 1500, respectively (\cite{DWE_lor98}).

The data were processed to a calibrated spectrum using the procedures and
calibration files of the ISO off-line pipeline version 7.0.
The individual sub-band spectra, when combined into
a single spectrum, can show jumps in flux levels at the band
edges. This is due to imperfect flux calibration or wrong
dark-current subtraction for low-flux observations. Using the
overlap regions of the different sub-bands and looking at other
SWS observations, several sub-bands were multiplied by a small
factor to construct a smooth spectrum. Note
that all shifts are well within the photometric absolute
calibration uncertainties claimed by \cite*{DWE_sch96} and
\cite*{DWE_feucht98}. 

\vspace*{-.3truecm}
\subsection{Literature study}

In the framework of this research, a detailed literature study was
indispensable. On the one hand, this was necessary to extract the best possible
set of starting fundamental parameters in order to reduce the number of
calculated spectra. On the other hand, one then could
check the consistency between the stellar parameters deduced from the ISO-SWS
data and other spectra/methods.

For all the selected stars, we therefore have scanned the
literature using SIMBAD (Set of Identifications, Measurements,
and Bibliography for Astronomical Data) and ADS (Astrophysics Data
System). An exhaustive discussion of published results is
presented in \cite*{DWE_decinthes}.

\vspace*{-.3truecm}
\subsection{Influence of stellar parameters}

In order to solve problems with the theoretical atmospheric structure, it is
important to know the relative importance of the different molecules and the
influence of the stellar parameters on the total absorption of the different
atoms and molecules. This is described in \cite*{DWE_decin97} and
\cite*{DWE_decin2000}).  

In spite of the moderate resolution of ISO-SWS, \cite*{DWE_decin2000} have
demonstrated that one can pin down the stellar parameters of the cool giants
very accurately from these data. This is due to the large wavelength range of
ISO-SWS, where different molecules determine the spectral signature. Since these
different features do each react in another way to a change of one of the
several heterogeneous stellar parameters (being the effective temperature, the
gravity, the mass, the metallicity, the microturbulent velocity, the \cc\ ratio
and the abundance of carbon, nitrogen and oxygen), it is possible to improve the
initial stellar parameters deduced from the literature
study. 

The method of analysis, which has been described in \cite*{DWE_decin2000}, could
however not be applied to the {\it{hot}} stars of the sample. Absorption by
atoms determines the spectrum of these stars. It turned out to be unfeasable to
determine the stellar parameters from the ISO-SWS spectra of these {\it{hot}}
stars, due to 1. problems with atomic oscillator strengths in the infrared (see
Sect. \ref{DWE_sec:res}); 2. the small dependence of the continuum on the
fundamental parameters (when changed within their uncertainty); 3. the small
dependence of the atomic line strength on the fundamental parameters; and 4.
the absence of molecules, which are each of them specifically dependent on the
various stellar parameters. Therefore, good-quality published stellar parameters
were used to compute the theoretical model and corresponding synthetic
spectrum. The angular diameter, stellar radius, mass and luminosity were then
calculated from the ISO-SWS spectrum.

\vspace*{-.3truecm}
\subsection{Statistical method}

Due to the high level of flux accuracy, a statistical method was
needed to evaluate the different synthetic spectra with respect to
each other. A choice was made for the Kol\-mo\-go\-rov-Smirnov test.
This well-developed goodness-of-fit criterion is applicable for a
broad range of comparisons between two samples, where the kind of
differences which occur between the samples can be very diverse.
An elaborate discussion about the Kolmogorov-Smirnov statistics
can be found in Pratt and Gibbons (1981) and H\'{a}jek (1969). 
How this statistical method can be applied for astronomical purposes, is
described in \cite*{DWE_decin2000}. The Kolmogorov-Smirnov
test {\it{globally}} checks the goodness of fit of the observed
and synthetic spectra. An advantage of this test is that one very
discrepant frequency point (e.g. due to a wrong oscillator
strength) only mildly influences the final result. Due to the smaller weights
which are given automatically to 
small features, the traditional comparison between observed and
synthetic spectra by eye-ball fitting is still necessary as a
complement to this Kolmogorov-Smirnov method in order to reveal
systematic errors in those features. The final error bars on the
atmospheric parameters are then estimated from 1. the intrinsic
uncertainty on the synthetic spectrum (i.e. the possibility to
distinguish different synthetic spectra at a specific resolution,
i.e. there should be a significant difference in the deviation estimating
parameter, $\beta$, calculated from the Kolmogorov-Smirnov test)
which is thus dependent on both the resolving power of the
observation and the specific values of the fundamental parameters,
2. the uncertainty on the ISO-SWS spectrum which is directly
related to the quality of the ISO-SWS observation , 3. the value
of the $\beta$-parameters in the Kolmogorov-Smirnov test and 4.
the still remaining discrepancies between observed and synthetic
spectrum. However, no exact formula can be given to compute the
error bars since the several parameters are not independent. 

\vspace*{-.3truecm}
\subsection{High-resolution observations}

To test our findings with data taken with an independent
instrument, a high-resolution observation of both one {\it{hot}}
and one {\it{cool}} star are studied very carefully. The
high-resolution Fourier Transform Spectrometer (FTS) spectrum of
$\alpha$ Boo (\cite{DWE_hinkle95}) and the ATMOS spectrum of the
Sun (\cite{DWE_farm89}, \cite{DWE_geller89}) are used as external
control to the process. 

\vspace*{-.3truecm}
\section{RESULTS}\label{DWE_sec:res}

\subsection{Stellar parameters}

Computing synthetic spectra is one step, distilling useful
information from it is a second - and far more difficult - one.
Fundamental stellar parameters for this sample of bright stars are
a first direct result which can be deduced from this comparison
between ISO-SWS data and synthetic spectra. These fundamental parameters
- deduced from the ISO-SWS spectrum for the {\it{cool}} stars and taken from the
literature for the {\it{hot}} stars - are summarized in Table
\ref{DWE_tab:stellparam}. In this table, the
effective temperature in K, the gravity in cm/s$^2$, the
microturbulent velocity in \km, the metallicity, the abundance of
carbon, nitrogen and oxygen, the isotopic ratio \cc\ and the
(spectrophotometric) angular diameter in mas are given as the
first ten parameters. From the parallax measurements (mas) of
Hipparcos (with an exception being $\alpha$ Cen A, for which a
more accurate parallax by \cite*{DWE_pour99} is available),
one may deduce the distance D (in pc). With the angular diameter
from ISO-SWS, the stellar radius R (in \Rsun) is calculated,
which, combined with the gravity, implies the gravity-induced mass
M$_g$ (in \Msun). The luminosity, extracted from the radius and
the effective temperature, is the last physical quantity listed in
this table. 

\begin{table*}[!ht]
\caption{\label{DWE_tab:stellparam} Final fundamental stellar parameters
for the selected stars in the sample. The effective temperature
\teff\ is given in K, the logarithm of the gravity in c.g.s.
units, the microturbulent velocity \vt\ in km/s, the angular
diameter in mas, the parallax $\pi$ in mas, the distance D in
parsec, the radius R in \Rsun, the gravity-induced mass M$_g$ in
\Msun\ and the luminosity L in \Lsun.}
\begin{center}
\leavevmode
\footnotesize 
\setlength{\tabcolsep}{1.2mm}
\begin{tabular}{|l||c|c|c|c|c|c|c|c|c|}  \hline
\rule[-3mm]{0mm}{8mm}  & \object{$\alpha$ Lyr} & \object{$\alpha$ Cma} &
\object{$\beta$ Leo} & \object{$\alpha$ Car} & \object{$\alpha$ Cen A} &
\object{$\delta$ Dra} & \object{$\xi$ Dra} & \object{$\alpha$ Boo} \\ 
\hline
\rule[0mm]{0mm}{5mm}Sp. Type & A0~V & A1~V & A3~Vv & F0~II
& G2~V & G9~III & K2~III & K2~IIIp \\ 
\teff & $9700 \pm 200$ & $10150 \pm 400$ & $8630
\pm 300$ & $7350 \pm 250$ & $5830 \pm 30$  & $4820 \pm 70$ & $4420 \pm 150$ &
$4300 \pm 70$ \\ 
$\log$ g & $3.95 \pm 0.20$ & $4.30 \pm 0.20$ & $4.20 \pm 0.25$ & $1.80 \pm
0.25$ & $4.35 \pm 0.05$ & $2.90 \pm 0.15$ & $2.60 \pm 0.15$ & $1.50 \pm 0.15$ \\
\vt & $2.0 \pm 0.5$ & $2.0 \pm 0.5$ & $2.0$ & $2.0$ & $1.0 \pm 0.1$ & $1.7 \pm
0.5$ & $2.0 \pm 1.0$ & $1.7 \pm 0.3$ \\
${\mathrm{[Fe/H]}}$ & $-0.40 \pm 0.30$ & $0.50 \pm 0.30$ & $0.00$ & $0.00$ &
$0.25 \pm 0.02$ & $0.00 \pm 0.25$ & $-0.10 \pm 0.30$ & $-0.50 \pm 0.20$ \\
$\varepsilon$(C) & $8.42 \pm 0.15$& $7.97 \pm 0.15$
& $8.56 \pm 0.20$ & $8.41 \pm 0.10$ & $8.74 \pm 0.05$ & $8.25 \pm 0.25$ & $8.26
\pm 0.30$ & $7.96 \pm 0.15$ \\
$\varepsilon$(N) & $8.00 \pm 0.15$ & $8.15 \pm 0.15$ & $8.05 \pm 0.20$ & $8.68
\pm 0.05$ & $8.26 \pm 0.09$ & $8.26 \pm 0.25$ & $8.26 \pm 0.30$ & $7.55 \pm
0.15$ \\ 
$\varepsilon$(O) & $8.74 \pm 0.15$  & $8.55 \pm 0.12$ & $8.93 \pm 0.20$ & $8.91
\pm 0.10$ & $9.13 \pm 0.06$ &  $8.83 \pm 0.25$ & $8.93 \pm 0.30$ & $8.67 \pm
0.15$ \\
\cc & $89$ & $89$ & 89 & 89 & 89 & $12 \pm 2$ & $20 \pm 5$ & $7 \pm 1$ \\ 
\ad & $3.35 \pm 0.16$ & $6.17 \pm 0.27$ & $1.47 \pm 0.06$
& $7.22 \pm 0.30$ & $8.80 \pm 0.344$ & $3.31 \pm 0.13$ & $3.09 \pm 0.12$ &
$20.80 \pm 0.83$ \\ 
$\pi$ & $128.93 \pm 0.55$ & $379.21 \pm 1.58$ & $90.6 \pm 0.89$ & $10.43 \pm
0.53$ & $737 \pm 2.6$ & $32.54 \pm 0.46$ & $29.26 \pm 0.49$ & $88.85 \pm 0.74$
\\  
D & $7.76 \pm 0.03$ & $2.63 \pm 0.01$ & $11.09 \pm 0.11$ & $95.88 \pm 4.87$ &
$1.36 \pm 0.01$  & $30.73 \pm 0.43$ & $34.34 \pm 0.57$ & $11.26 \pm 0.09$ \\ 
R & $2.79 \pm 0.13$ & $1.75 \pm 0.08$ & $1.75 \pm 0.07$ & $74.39 \pm
4.89$ & $1.27 \pm 0.05$ & $10.96 \pm 0.46$ & $11.28 \pm 0.48$ & $25.24 \pm 1.03$
\\  
M$_g$ & $2.54 \pm 1.19$ & $2.23 \pm 1.05$  & $1.77 \pm 0.82$  & $12.8 \pm 6.13$
& $1.30 \pm 0.46$ & $2.77 \pm 0.98$ & $1.87 \pm 1.30$ & $0.74 \pm 0.26$ \\  
\rule[-3mm]{0mm}{3mm}L & $62 \pm 5$ & $29
\pm 4$ & $15\pm 2$ & $14571 \pm 497$ & $1.7 \pm 0.2$ & $56 \pm 6$ & $44 \pm 7$
& $196 \pm 21$ \\ \hline 
\end{tabular}

\hspace*{.1cm}\begin{tabular}{|l||c|c|c|c|c|c|c|c|}  \hline
\rule[-3mm]{0mm}{8mm} & \object{$\alpha$ Tuc} & \object{$\beta$ UMi} &
\object{$\gamma$ Dra} & \object{$\alpha$ Tau} & \object{HD~149447} &
\object{$\beta$ And} & \object{$\alpha$ Cet} & \object{$\beta$ Peg}
\\ \hline \rule[0mm]{0mm}{5mm}Sp. Type & K3~III & K4~III & K5~III
&  K5~III & K6~III & M0~III & M2~III & M2.5~III\\
\teff & $4040 \pm 70$ & $4150 \pm 70$ & $3930 \pm 70$ & $3850 \pm
70$ & $3900 \pm 70$ & $3780 \pm 70$ & $3745 \pm 70$ & $3590 \pm 150$
\\ $\log$ g & $1.10 \pm 0.15$ & $ 1.90 \pm 0.15$ & $1.55 \pm
0.25$ & $1.50 \pm 0.15$ & $1.10 \pm 0.15$ & $1.40 \pm 0.20$ & $1.30 \pm 0.15$ &
$1.50 \pm 0.40$ 
\\ \vt & $1.7 \pm 0.5$ & $2.0 \pm 0.5$ & $2.0 \pm 0.5$
& $1.7 \pm 0.3$ & $2.0 \pm 0.3$ & $2.3 \pm 0.5$ & $2.3 \pm 0.5$ & $2.3 \pm
 0.5$ \\
${\mathrm{[Fe/H]}}$ & $0.00 \pm 0.20$ & $-0.15 \pm 0.20$ &
$0.00 \pm 0.20$ & $-0.15 \pm 0.20$ & $0.00 \pm 0.20$ & $0.00 \pm 0.20$ &
 $0.00 \pm 0.20$ & $0.00 \pm 0.30$
\\ $\varepsilon$(C) & $8.24 \pm 0.20$ & $8.40 \pm 0.20$ &
$8.22 \pm 0.20$ & $8.35 \pm 0.20$ & $8.16 \pm 0.20$ & $8.20 \pm 0.20$ &
$8.40 \pm 0.20$ & $8.56 \pm 0.30$ \\ 
$\varepsilon$(N) & $8.26 \pm 0.20$ & $8.16 \pm 0.20$ & $8.26
\pm 0.20$ & $8.35 \pm 0.20$ & $8.26 \pm 0.20$ & $8.37 \pm 0.20$ & $8.26 \pm
0.20$ & $8.24 \pm 0.30$
\\ $\varepsilon$(O) & $8.73 \pm 0.15$ & $8.83 \pm 0.15$ &
$8.81 \pm 0.15$ & $8.83 \pm 0.15$ & $9.03 \pm 0.15$ & $8.84 \pm 0.15$ &
$8.93 \pm 0.15$ & $9.03 \pm 0.30$ \\ 
\cc & $23 \pm 3$ & $9 \pm 1$ & $10 \pm 1$ & $10 \pm 1 $ & $8 \pm 1$
& $ 9 \pm 1$ & $10 \pm 1$ & $7 \pm 1$ \\
\ad & $6.23 \pm 0.25$ & $9.86 \pm 0.40$ & $10.07 \pm 0.40$
& $20.77 \pm 0.83$ & $4.75 \pm 0.19$ & $13.59 \pm 0.55$ & $12.52 \pm 0.50$ &
$16.88 \pm 0.70$ 
\\ $\pi$ & $16.42 \pm 0.59$ & $25.79 \pm 0.52$ & $22.10 \pm
0.46$ & $20.77 \pm 0.83$ & $4.75 \pm 0.19$ & $13.59 \pm 0.55$ & $12.52 \pm 0.50$
& $16.88 \pm 0.70$ 
\\ D & $60.90 \pm 2.19$ & $38.78 \pm 0.78$ & $45.25 \pm
0.94$ & $19.96 \pm 0.38$ & $103.84 \pm 8.52$ & $61.12 \pm 2.84$ & $67.48 \pm
3.78$ & $61.08 \pm 2.69$
\\ R & $40.77 \pm 2.18$ & $41.09 \pm 1.87$ & $48.97 \pm 2.21$
& $44.63 \pm 1.97$ & $52.95 \pm 4.81$ & $89.27 \pm 5.50$ &
$90.86 \pm 6.25$ & $110.52 \pm 6.69$
\\ M$_g$ & $0.76 \pm 0.27$ & $4.90 \pm 1.80$ & $3.11 \pm 1.50$
& $2.30 \pm 0.82$ & $1.29 \pm 0.45$ & $7.32 \pm 3.48$ &
$6.02 \pm 2.89$ & $14.12 \pm 6.73$ \\ 
\rule[-3mm]{0mm}{3mm}L & $393 \pm 50$ & $452 \pm 48$ & $516 \pm
59$ & $395 \pm 45$ & $577 \pm 113$ & $1468 \pm 211$ & $1465 \pm
 229$ & $1830 \pm 300$\\ \hline 
\end{tabular}
\end{center}
\end{table*}

\vspace*{-.3truecm}
\subsection{Discussion on discrepancies}\label{DWE_subsec:disc}

The discrepancies between the ISO-SWS and synthetic spectra are subjected to a
careful scrutiny in order to elucidate their origin. 
A typical example of both a
{\it{hot}} and {\it{cool}} star is given in Fig. \ref{DWE_fig:acar} and
Fig. \ref{DWE_fig:atau} respectively. 
A description on the general trends in the discrepancies for {\it{hot}} and
{\it{cool}} stars is given. 

\vspace*{-.3truecm}
\subsubsection{{\it{Hot}} stars: A0-G2}\label{DWE_subsubsec:hot}

\begin{enumerate}
\item{When concentrating e.g. on $\alpha$ Cen A (G2 V),
one notifies quite a few spectral features which appear in the
ISO-SWS spectrum, but are absent in the synthetic spectrum. Some
of the most prominent ones are indicated by an arrow in Fig.
\ref{DWE_fig:acenarrow}. The solar ATMOS spectrum proved to be extremely
useful for the determination of the origin of these features. All
spectral features, indicated by an arrow in Fig. \ref{DWE_fig:acenarrow},
turned out to be caused by - strong - atomic lines (Mg, Si, Fe, Al,
C, ...). The usage of other atomic line lists did not solve the problem.

\begin{figure}[h!]
\begin{center}
\resizebox{3.4in}{!}{\rotatebox{90}{\includegraphics{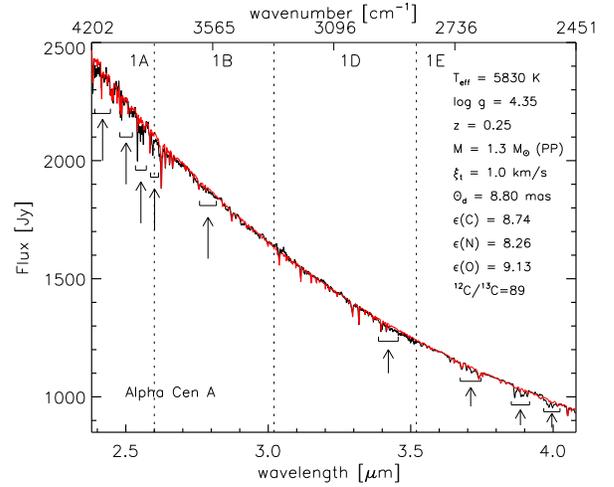}}}\vspace{-.5truecm}
\caption{\label{DWE_fig:acenarrow} Comparison between the ISO-SWS data of
$\alpha$ Cen A (black) and the synthetic spectrum (red) with
stellar parameters \teff\ = 5830 K, $\log$ g = 4.35, M = 1.3
\Msun, z = 0.25, \vt\ = 1.0 \km, \cc\ = 89, $\varepsilon$(C) =
8.74, $\varepsilon$(N) = 8.26, $\varepsilon$(O) = 9.13 and \ad\ =
8.80 mas. Some of the most prominent discrepancies between these
two spectra are indicated by an arrow.}
\end{center}
\end{figure}

The lack of reliable atomic data rendered the determination of the continuum of
the {\it{hot}} stars
very difficult. As a consequence, the uncertainty on the angular diameter is
more pronounced. Therefore, Vega and Sirius have been used to check our
findings, but they have not been studied into all detail. In order to deduce
possible problems with calibration files, more trustworthy data as input for the
theoretical models are needed. }

\item{The hydrogen lines are also conspicuous. For example,
the synthetic hydrogen Pfund lines are almost always
predicted as too strong for main-sequence stars, while they are
predicted as too weak for the supergiant $\alpha$ Car (see Fig.
\ref{DWE_fig:acar}, where the hydrogen lines are indicated by an arrow).
This indicates a problem with the generation of the synthetic
hydrogen lines, which is corroborated when the high-resolution
ATMOS spectrum of the Sun is compared with its synthetic spectrum.

\vspace*{-.5truecm}
\begin{figure}[h!]
\begin{center}
\resizebox{3.4in}{!}{\rotatebox{90}{\includegraphics{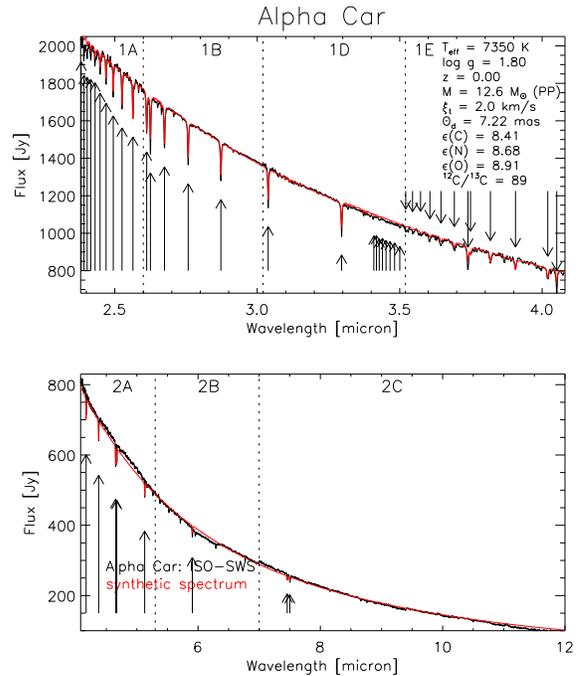}}}\vspace{-.5truecm}
\caption{\label{DWE_fig:acar} Comparison between band 1 and band 2 of the
ISO-SWS data of $\alpha$ Car (black) and the synthetic spectrum
(red) with stellar parameters \teff\ = 7350 K, $\log$ g = 1.80, M
= 12.6 \Msun, z = 0.00, \vt\ = 2.0 \km, \cc\ = 89,
$\varepsilon$(C) = 8.41, $\varepsilon$(N) = 8.68, $\varepsilon$(O)
= 8.91 and \ad\ = 7.22 mas. Hydrogen lines are indicated by an
arrow.}
\end{center}
\end{figure}
}

\item{Compared to the ISO-SWS data, the synthetic spectra of {\it{hot}} stars
display a higher flux between the H5-9 and H5-8 hydrogen
line (see Fig. \ref{DWE_fig:acar}). From other SWS observations
available in the ISO data-archive, we could deduce that this
'pseudo-continuum' starts arising for stars hotter than K0 ($\sim
4500$ K). Since such an effect was not seen for the cooler K and M
giants, we could reduce the problem as having an atomic
origin. A scrutiny on the hydrogen lines learns that the
high-excitation Humphreys-lines (from H6-18 on) - and Pfund-lines -
are always calculated as too weak. Moreover, the Humphreys
ionization edge occurs at 3.2823 \mic. Since the discrepancy does
not appear above the limit (i.e. at shorter wavelengths) and is
disappearing beyond the Brackett-$\alpha$ line, the conclusion is
reached that the missing lines could well be the crowding of
Humphreys hydrogen lines towards the series limit.}

\item{From 3.84 \mic\ on, fringes at the end of band 1D affect the
SWS spectrum of almost all stars in the sample.}

\item{A clear discrepancy is visible at the beginning of band 1A.
For the {\it{hot}} stars, the H5-22 and H5-23 lines emerge in that
part of the spectrum. An analogous discrepancy is also seen for
the {\it{cool}} stars, though it is somewhat more difficult to
recognize due to the presence of many CO features (Fig.
\ref{DWE_fig:atau}). Being present in the continuum of both {\it{hot}}
and {\it{cool}} stars, this discrepancy is attributed to problems
with the RSRF. A broad-band correction was already applied at the
short-wavelength edge of band 1A (\cite{DWE_vdb99}), but the
problem seems not to be fully removed. At the band edges, the
responsivity of a detector is always small. Since the data are
divided by the RSRF, a  small problem with the RSRF at these
places may introduce a pronounced error at the band edge.}

\item{Memory effects make the calibration of band 2 for all the stars very
difficult. These memory effects are more severe for the
{\it{cool}} stars, since the CO and SiO absorptions cause a steep
increase (decrease) in flux for the up (down) scan. The RSRFs for the
sub-bands will therefore only be modeled well once there is a
full-proof method to correct SWS data for detector memory effects.}
\end{enumerate}

\vspace*{-.6truecm}
\subsubsection{{\it{Cool}} stars: G2-M2}\label{DWE_subsubsec:cool}

\begin{enumerate}
\item{The situation changes completely when going to the {\it{cool}} stars of
the sample. While the spectrum of the {\it{hot}} stars is
dominated by atomic-line features, molecules determine the
spectral signature of the {\it{cool}} stars (Fig. \ref{DWE_fig:atau}).  A few of
the - problematic - atomic features (see Sect. \ref{DWE_subsubsec:hot}) can
still be identified in these cool stars. E.g. atomic spectral
feature around 3.97 \mic\ (Fig. \ref{DWE_fig:acenarrow}) remains visible for
the whole sample, even till $\alpha$ Cet.}
\vspace*{-.5truecm}
\begin{figure}[!h]
\begin{center}
\resizebox{3.4in}{!}{\rotatebox{90}{\includegraphics{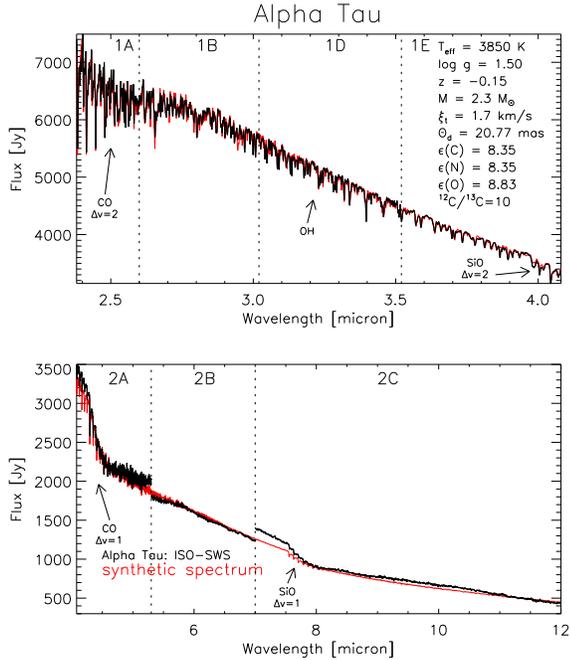}}}\vspace{-.5truecm}
\caption{\label{DWE_fig:atau} Comparison between band 1 and band 2 of the
ISO-SWS data of $\alpha$ Tau (black) and the synthetic spectrum
(red) with stellar parameters \teff\ = 3850 K, $\log$ g = 1.50, M
= 2.3 \Msun, z = -0.15, \vt\ = 1.7 \km, \cc\ = 10,
$\varepsilon$(C) = 8.35, $\varepsilon$(N) = 8.35, $\varepsilon$(O)
= 8.83 and \ad\ = 20.77 mas. The most important absorbers are
indicated by an arrow.}
\end{center}
\end{figure}

\item{\vspace*{-.6truecm}One of the most prominent molecular features in band 1
is the first-overtone band of carbon monoxide (CO, $\Delta v =2$)
around 2.4 \mic. In these oxygen-rich stars, the total amount of carbon
determines the strength of the CO band. This carbon abundance can be
calculated in two ways:
\begin{itemize}
\item{the strong CO absorption causes a 'dip' in the continuum
till $\sim$ 2.8 \mic, which may be used to determine
$\varepsilon$(C);}
\item{the strength of the CO spectral features is directly related
to $\varepsilon$(C).}
\end{itemize}
Computing a synthetic spectrum with the carbon abundance
determined from this first criterion, results however in the
(strongest) CO spectral features being always too strong compared to the
ISO-SWS observation (2-4\%). It has to be noted that this mismatch
occurs in band 1A, where the standard deviation of the rebinned
spectrum is larger than for the other sub-bands and that the error is within the
quoted accuracy of 
ISO-SWS in band 1A (\cite{DWE_sch96}). Nevertheless, it is
alarming that this mismatch is not random, in the sense that the
observed CO features are {\it{always}} weaker than the synthetic
ones.

The agreement between the high-resolution FTS spectrum and synthetic spectrum of
$\alpha$ Boo is however extremely good. When scrutinizing carefully the
first-overtone CO lines in the FTS spectrum, it was clear that
all the $^{12}$CO 2-0 lines, and practically all the $^{12}$CO 3-1
lines, are predicted as too {\it{weak}} (by 1-2\%)! 

Firstly, it has to be noted that the flux values in the
wavelength region from 2.38 to 2.4 \mic, where the CO 2-0 P18 and
the CO 2-0 P21 are the main features, are unreliable due to
problems with the RSRF of band 1A. Secondly, no correlation is found with the
local minima and maxima in the RSRF of band 1A. Since the instrumental
profile of an AOT01 is still not exactly known, the synthetic data were
convolved with a gaussian with FWHM=$\lambda$/resolution. This incorrect
gaussian instrumental profile introduces an error which will be most visible on
the strongest lines. Together with too high a theoretical resolving
power for an AOT01 speed-4 observation in band 1A ($\ge 1500$), these two
instrumental effects may explain this discrepancy.}

\item{For both the high-resolution FTS
and the medium-resolution SWS spectrum (Fig. \ref{DWE_fig:atau}), the strongest
lines (OH 1-0 lines) are predicted as 
too weak, while the other lines match very well.
Since the same effect occurs for these two different observations,
it it plausible to assume that the origin of the problem is
situated in the theoretical model or in the synthetic-spectrum
computation. Wrong oscillator strengths for the OH lines could
e.g. cause that kind of problems. Different OH line lists do however all show
the same trend (\cite{DWE_decinthes}). Since a similar 
discrepancy was also noted for the low-excitation CO lines, a problem with the
temperature distribution in the outermost layers of the stellar
model is a very plausible explanation for this discrepancy.}

\item{The same remarks as for the {\it{hot}} stars, concerning the fringes and the memory effects, can be given.}

\end{enumerate}

So far, the origins of all the general discrepancies in band 1 for both
{\it{hot}} and {\it{cool}} stars have been traced. With this in mind, the other
stars of our sample have been studied carefully and the results are described in
\cite*{DWE_decinthes}.

\vspace*{-.3truecm}
\section{IMPLICATION ON CALIBRATION AND MODELING}\label{DWE_sec:impact}
The results of this detailed comparison between observed ISO-SWS data and
synthetic spectra have an impact both on the calibration of the ISO-SWS data and
on the theoretical description of stellar atmospheres.

From the calibration point of view, a first conclusion is
reached that the broad-band shape of the relative spectral response function is
at the moment already quite accurate, although some improvements
can be made at the beginning of band 1A and band 2. Also, a fringe pattern is
recognized at the 
end of band 1D. Inaccurate beam profiles together with too high a resolving
power may cause the strongest CO lines to be predicted as too strong. Since the
same molecules are absorbing in band 1 and in band 2, these synthetic
spectra are supposed to be also very accurate in band 2. These
spectra will therefore be used to test the recently developed
method for memory effect correction (\cite{DWE_kester2000}) and to rederive
the relative spectral response function for band 2. The synthetic
spectra of the standard sources of our sample are not only used to
improve the flux calibration of the observations taken during the
nominal phase, but they are also an excellent tool to characterize
instabilities of the SWS spectrometers during the post-helium
mission.

Concerning the modeling part, problems with the construction of
the theoretical model and computing of the synthetic spectra are
pointed out. The comparison between the high-resolution FTS
spectrum of $\alpha$ Boo and the corresponding synthetic spectrum
revealed that the low-exci\-ta\-tion first-overtone CO lines and
fundamental OH lines are predicted as too weak. This indicates a
problematic temperature distribution in the outermost layers of
the theoretical models. The upper photosphere is very difficult
to model and is often computed from an extrapolation of the
interior layers. The temperature distribution should now be
disturbed in order to simulate a chromosphere, convection, a
change in opacity, ... Using these improved models, the change in
abundance pattern resulting from the ISO-SWS data should be
studied.
The complex computation of the hydrogen lines, together with the
inaccurate atomic oscillator strengths in the infrared rendered
the computation of the synthetic spectra for {\it{hot}} stars
difficult. In spite of the fact that the broadening parameters
were thought to be taken into account properly, the ISO-SWS data
displayed a notorious discrepancy for the Brackett, Pfund and
Humphreys lines. People from the stellar-atmosphere-group in
Uppsala are now scrutinizing this problem. The high-resolution
ATMOS spectrum of the Sun and the SWS spectra of our standard
sources indicated an insufficient knowledge of atomic oscillator
strengths in the infrared. J. Sauval (Royal Observatory Belgium)
is now trying to derive empirical oscillator strengths from the
high-resolution ATMOS spectrum of the Sun.

Although we have mainly concentrated on the discrepancies between
the ISO-SWS and synthetic spectra -s ince this was the main task of
this research - we would like to emphasize the very good agreement
between observed ISO-SWS data and theoretical spectra. The very
small discrepancies still remnant in band 1 are at the 1-2\%
level for the giants, proving not only that the calibration of the (high-flux)
sources has already reached a good level of accuracy, but also
that the description of cool star atmospheres and molecular line
lists is very accurate. The theoretical description of cool star
atmospheres has lagged behind for a long time the description of
hot star atmospheres, but, I think, that from now on, we may say
that this is not anymore true!


\vspace{-.5truecm}
\begin{thebibliography}{}

\bibitem[\protect\astroncite{Cohen et~al.}{1992}]{DWE_cohen92}
Cohen M., Walker R.G., Witteborn F.C., 1992, AJ 104,
2030

\bibitem[\protect\astroncite{Cohen et~al.}{1995}]{DWE_cohen95}
Cohen M., Witteborn F.C., Walker R.G., Bregman J., Wooden
D.H., 1995, AJ 110, 275

\bibitem[\protect\astroncite{Cohen et~al.}{1996b}]{DWE_cohen96b}
Cohen M., Witteborn F.C., Carbon D.F., Davies J.K., Wooden D.H., Bregman
J.D., 1996b, AJ 112, 2274

\bibitem[\protect\astroncite{Decin et~al.}{1997}]{DWE_decin97}
Decin L., Cohen M., Eriksson K., Gustafsson B., Huygen E., Morris P.,
 Plez B., Sauval J., Vandenbussche B., Waelkens C., 1997, in `Proc. First
ISO Workshop on  Analytical Spectroscopy',  Heras, A.M., Leech, K., Trams,
N.R., Perry,  M. (eds.), ESA-SP 419, p. 185 

\bibitem[\protect\astroncite{Decin et~al.}{2000}]{DWE_decin2000}
Decin L., Waelkens C., Eriksson K., Gustafsson B., Plez B., Sauval A.J., Van
Assche W., Vandenbussche B., 2000, A\&A, submitted

\bibitem[\protect\astroncite{Decin}{2000}]{DWE_decinthes}
Decin L., 2000, in `Synthetic spectra of cool stars observed with the ISO
Short-Wavelength Spectrometer: improving the models and the calibration of the
instrument', PhD. thesis, University of Leuven 

\bibitem[\protect\astroncite{Farmer \& Norton}{1989}]{DWE_farm89}
Farmer C.B., Norton R.H., 1989, in `Atlas of the Infrared
Spectrum of the Sun and the Earth Atmosphere from Space. Volume I,
The Sun', NASA Reference Publication 1224

\bibitem[\protect\astroncite{Feuchtgruber}{1998}]{DWE_feucht98}
Feuchtgruber H., 1998, in `Status of AOT01 AOT band order ratios
and related data', ISO-SWS online documentation

\bibitem[\protect\astroncite{Geller}{1989}]{DWE_geller89}
Geller M., 1989, in `Atlas of the Infrared Spectrum of the
Sun and the Earth Atmosphere from Space. Volume III. Key to
Identification of Solar Features', NASA Reference Publication
1224

\bibitem[\protect\astroncite{Gustafsson et~al.}{1975}]{DWE_gust75}
Gustafsson B., Bell R.A., Eriksson K., Nordlund
{\AA}, 1975, A\&A 42, 407

\bibitem[\protect\astroncite{Hinkle et~al.}{1995}]{DWE_hinkle95}
Hinkle K., Wallace L., Livingston W., 1995, PASP 107, 1042

\bibitem[\protect\astroncite{Kester}{2000}]{DWE_kester2000}
Kester D., 2000, in `ISO beyond the Peaks. The 2nd ISO
workshop on analytical spectroscopy.', A. Salama (eds.), SP-456,
p. 93

\bibitem[\protect\astroncite{Lorente}{1998}]{DWE_lor98}
Lorente R., 1998, in `Spectral Resolution of SWS AOT 1', ISO-SWS
online documentation

\bibitem[\protect\astroncite{Monnier et~al.}{1998}]{DWE_mon98}
Monnier J.D., Geballe T.R., Danchi W.C., 1998, ApJ 502, 833

\bibitem[\protect\astroncite{Plez et~al.}{1992}]{DWE_plez92}
Plez B., Brett J.M., Nordlund {\AA}, 1992,
A\&A 256, 551

\bibitem[\protect\astroncite{Plez et~al.}{1993}]{DWE_plez93}
Plez B., Smith. V.V., Lambert D.L., 1993, AJ 418, 812

\bibitem[\protect\astroncite{Pourbaix et~al.}{1999}]{DWE_pour99}
Pourbaix D., Neuforge-Verheecke C., Noels A., 1999, A\&A 344, 172

\bibitem[\protect\astroncite{Schaeidt et~al.}{1996}]{DWE_sch96}
Schaeidt S.G., Morris P.W., Salama A., et al., 1996, A\&A 315, L55

\bibitem[\protect\astroncite{Tsuji et~al.}{1997}]{DWE_tsuji97}
Tsuji T., Ohnaka K., Aoki W., Yamamura I., 1997, A\&A 320, L1

\bibitem[\protect\astroncite{Vandenbussche}{1999}]{DWE_vdb99}
Vandenbussche B., 1999, in `The ISO-SWS Relative Spectral
Response Calibration', ISO-SWS online documentation

\bibitem[\protect\astroncite{Witteborn et~al.}{1999}]{DWE_witteborn99}
Witteborn F.C., Cohen M., Bregman J.D., Wooden D.H., Heere
K., Shirley E.L., 1999, AJ 117, 2552



\end{thebibliography}
\end{document}